\documentclass[sigconf]{acmart}
\AtBeginDocument{%
  }

\setcopyright{none}
\copyrightyear{}
\acmYear{}
\acmDOI{}
\acmConference[arXiv Preprint]{}{January,
  2026}{}
\acmISBN{}

\usepackage[normalem]{ulem} 

\acmSubmissionID{artpl\_230s1}

\begin{document}

%%\title{Revealing Fractal Pattern of Cellular Automata Through Recursive Gradient Profile Function}
\title{Revealing Latent Self-Similarity in Cellular Automata via Recursive Gradient Profiling}


\author{Chung-En Hao}
\affiliation{%
  \institution{Institute of Applied Arts, \\ National Yang Ming Chiao Tung Universit,}
  \city{Hsinchu}
  \country{Taiwan}
}
\email{alson6101@gmail.com}

\author{Ivan C. H. Liu}
\affiliation{%
  \institution{Institute of Applied Arts,  \\National Yang Ming Chiao Tung University,}
  \city{Hsinchu}
  \country{Taiwan}
}
\email{ivanliu@nycu.edu.tw}

\renewcommand{\shortauthors}{Hao, Liu}

%%\begin{abstract}
%%  Cellular automata (CA), initially developed as computational models of natural processes, have since become a central subject in complex systems research. Among them, the Ulam-Warburton Cellular Automaton (UWCA) displays recursive growth and fractal-like patterns in its spatial form. While exact fractal structures appear only at specific generations, this study introduces a Recursive Gradient Profile Function that maps grayscale values to each generation’s newly born cells, visually revealing latent self-similar fractal structures. We further explore UWCA variants using alternative neighborhood rules that also produce distinct fractal visuals. The underlying geometric logic of these generative patterns resonates with optical effects such as infinity mirror, video feedback, \textit{mise en abyme} in European art history, and fractal characteristics found in religious structures, revealing a potential connection between complexity science and cultural art and design.
%%\end{abstract}

\begin{abstract}
Cellular automata (CA), originally developed as computational models of natural processes, have become a central subject in the study of complex systems and generative visual forms. Among them, the Ulam–Warburton Cellular Automaton (UWCA) exhibits recursive growth and fractal-like characteristics in its spatial evolution. However, exact self-similar fractal structures are typically observable only at specific generations and remain visually obscured in conventional binary renderings. This study introduces a Recursive Gradient Profile Function (RGPF) that assigns grayscale values to newly activated cells according to their generation index, enabling latent self-similar structures to emerge cumulatively in spatial visualizations. Through this gradient-based mapping, recursive geometric patterns become perceptible across scales, revealing fractal properties that are not apparent in standard representations. We further extend this approach to UWCA variants with alternative neighborhood configurations, demonstrating that these rules also produce distinct yet consistently fractal visual patterns when visualized using recursive gradient profile. Beyond computational analysis, the resulting generative forms resonate with optical and cultural phenomena such as infinity mirrors, video feedback, and  \textit{mise en abyme} in European art history, as well as fractal motifs found in religious architecture. These visual correspondences suggest a broader connection between complexity science, computational visualization, and cultural art and design.
\end{abstract}




% 發表日期 等投稿上了才能更新

\maketitle

\section{Introduction}
From natural phenomena such as tree branching and snowflake crystallization to mathematical structures like Pascal’s triangle modulo 2 ~\cite{pascalTriangle_mod2}, the Sierpiński triangle, and the Koch snowflake, many complex forms emerge from simple, recursive rules. These self-similar fractal structures highlight a deep connection between equationcal logic and nature, forming a central focus in the study of complex systems \cite{fractal_book_Bunde, fractal_book_Peitgen, fractal_book_Falconer}.

Cellular Automata (CA), originally proposed by von Neumann and Ulam during early 1950s \cite{Schiff_CA}, which conceptualized to simulate biological growth and evolution using discrete systems . CA consists of cells arranged in a grid, with each cell updating its state based on local interactions. Among them, Conway’s Game of Life \cite{Conway_game} is the most widely known model. More broadly, the CA framework has been widely applied across fields including computer science, physics, earth science, social sciences, and even generative art \cite{Wolfram_new}.

The Ulam-Warburton Cellular Automaton (UWCA) was introduced by Ulam \cite{Ulam_uwca}, inspired by the growth patterns of crystals and plants. This model demonstrates fractal-like expansion \cite{Applegate_tooth} and infinite topological entropy, which indicate a high degree of complexity \cite{Kawaharada}. Later studies explored UWCA in modeling natural systems such as branching structures \cite{Okura_tree} and landscape formation \cite{Batty_geo}, and it has been recognized as the earliest model to simulate tree growth computationally \cite{Palubicki_tree}.

 Wolfram systematically studied both one-dimensional elementary CA and two-dimensional versions with binary states (live/dead), introducing a numbering system to analysis and classify their behavior. Many of these rules produce self-similar fractals in space-time diagrams, yet there has been little discussion of self-similar fractal patterns solely on spatial diagrams \cite{Wolfram_new}.

Among various CA, the UWCA stands out for its distinctive growth behavior suggesting a recursive growth. Applegate and co-wokers view its geometric shape and numerical progression as recursive, yet not truly self-similar fractal pattern \cite{Applegate_tooth}, others like Kawaharada have used the box-counting method to analysis its fractal dimension, fractal characteristics on the pattern outline with specific generations of evolution \cite{Kawaharada}.

Building on Applegate and co-workers' notion of recursion both geometrically and arithmetically, this work introduces Recursive Gradient Profile Function (RGPF), that applies grayscale profile to newly born cells in each generation. This visual mapping reveals self-similarity in the UWCA’s spatial structure that could not be captured through traditional black-and-white visualization. We also explore variant neighborhood configurations not previously discussed in the UWCA studies, and observe that these too produce self-similar fractal patterns by multiplying RGPF.

Finally, based on the properties of self-similarity and scale invariance in those UWCA images, we draw parallels between these visual forms and optical phenomena such as infinity mirrors ~\cite{infinityMirror}, video feedback ~\cite{fractal_book_Peitgen}, and \textit{mise en abyme} in European art history (refers to the concept “paintings within paintings”) ~\cite{mise}. We also relate them to fractal elements found in religious architecture and artifacts ~\cite{hindu_fractal, hindu_fractal2}, revealing a potential connection between complex systems, computational visualization, art and humanities.

\section{Background and Related Works}
This section introduce fractal geometry and its cultural interpretations, provide an overview of 1D and 2D cellular automata, and present key characteristics of UWCA.

\subsection{Fractal Geometry and Cultural Interpretations}
The term fractal refers to geometric structures that exhibit self-similarity across different scales and possess a fractal dimension~\cite{fractal_book_Falconer}. Self-similarity implies that a pattern is exactly or approximately similar to a part of itself, and the degree of this similarity across scales is expressed through fractal dimension. This dimension typically exceeds the object's topological dimension and is often non-integer~\cite{Cannon_fractal}. Moreover, a shape with non-integer fractal dimension is always fractal. The Hausdorff dimension, the most-used representation of fractal dimension, of a single point is zero, of a line segment is 1, of a square is 2, and of a cube is 3.

While fractal dimension can be rigorously defined in mathematics, natural systems rarely follow such ideal conditions. Noise, irregularity, and randomness in real-world data make direct application of Hausdorff's formula impractical, especially in image-based analysis. Nonetheless, the notion of fractal dimension remains a useful tool for analyzing and describing natural forms~\cite{fractal_book_Falconer}.

To address this, the box-counting method has become a widely adopted technique for estimating fractal dimensions in digital images~\cite{Firiytab_fractalDimension}. This approach involves converting an image to binary form, applying edge detection, and overlaying a grid of boxes to count how many contain part of the object’s boundary. Repeating this process at various scales reveals how the number of occupied boxes $N(\epsilon)$ scales with box size $\epsilon$, approximating the fractal dimension $D$:

\begin{equation}
D = \lim_{\varepsilon \to 0} \frac{\log N(\varepsilon)}{\log(1/\varepsilon)} .
\label{eq:fractal_dimension}
\end{equation}

In practice, this is estimated using a log–log plot of $\log N(\epsilon)$ versus $\log (1/\epsilon)$, with the slope of the best-fit line obtained via least squares regression.

Traditional box-counting methods are primarily used to estimate fractal dimension based on edges, such as in neuron structures or tree branches, where the outline of the object carries fractal properties. However, in cases where the fractal features are embedded in the grayscale texture itself (e.g., rock surfaces or Brownian motion), the edge-based method becomes insufficient. To overcome this limitation, Differential Box Counting (DBC) was introduced for analyzing grayscale images~\cite{Sarkar_DBC}.

To further improve accuracy, enhanced methods like Smoothed Differential Box Counting (SDBC) have been developed~\cite{Chen_SDBC}. The SDBC algorithm addresses common issues in traditional DBC, such as overcounting and sensitivity to gray-level shifts, by introducing a quantization and shifting mechanism along the intensity axis. In this study, we adopt SDBC to evaluate the fractal dimension of grayscale outputs  by our images.

\subsubsection{Fractal in Humanities and Art}
Beyond mathematics and nature, fractals also appear in artistic expression. The infinity mirror is a setup of parallel or angled mirrors arranged to produce reflections that seem to recede infinitely~\cite{infinityMirror}. A related media technique is video feedback~\cite{fractal_book_Peitgen}, where a camera captures its own output on a screen, creating recursive, real-time imagery that compresses toward the center. This visual effect resembles the concept of \textit{mise en abyme} in Western art history~\cite{Watling_Mise}, a technique involving images nested within themselves to suggest infinite regression—famously exemplified by Diego Velázquez’s painting \textit{Las Meninas}.

In many religious architectures, the use of repeated archways, in churches, mosques, or temples, creates a powerful visual rhythm that evokes a sense of recursion and infinity ~\cite{hindu_fractal}. As arches extend seemingly without end, one behind another, where each arch reflects and frames the next, the structure invites contemplation on the eternal, the divine, and the boundless nature of faith. This recursive form, as Acosta \cite{Acosta_recursive} suggests, mirrors the very architecture of reality itself---a reality not made of static things but of dynamic, self-updating processes. Religious architecture, then, becomes a material metaphor for this unfolding: a spatial expression of divine recursion.

\subsection{Cellular Automata and Self-Similarity}

\begin{figure}
  \centering
  \includegraphics[width=0.98\linewidth]{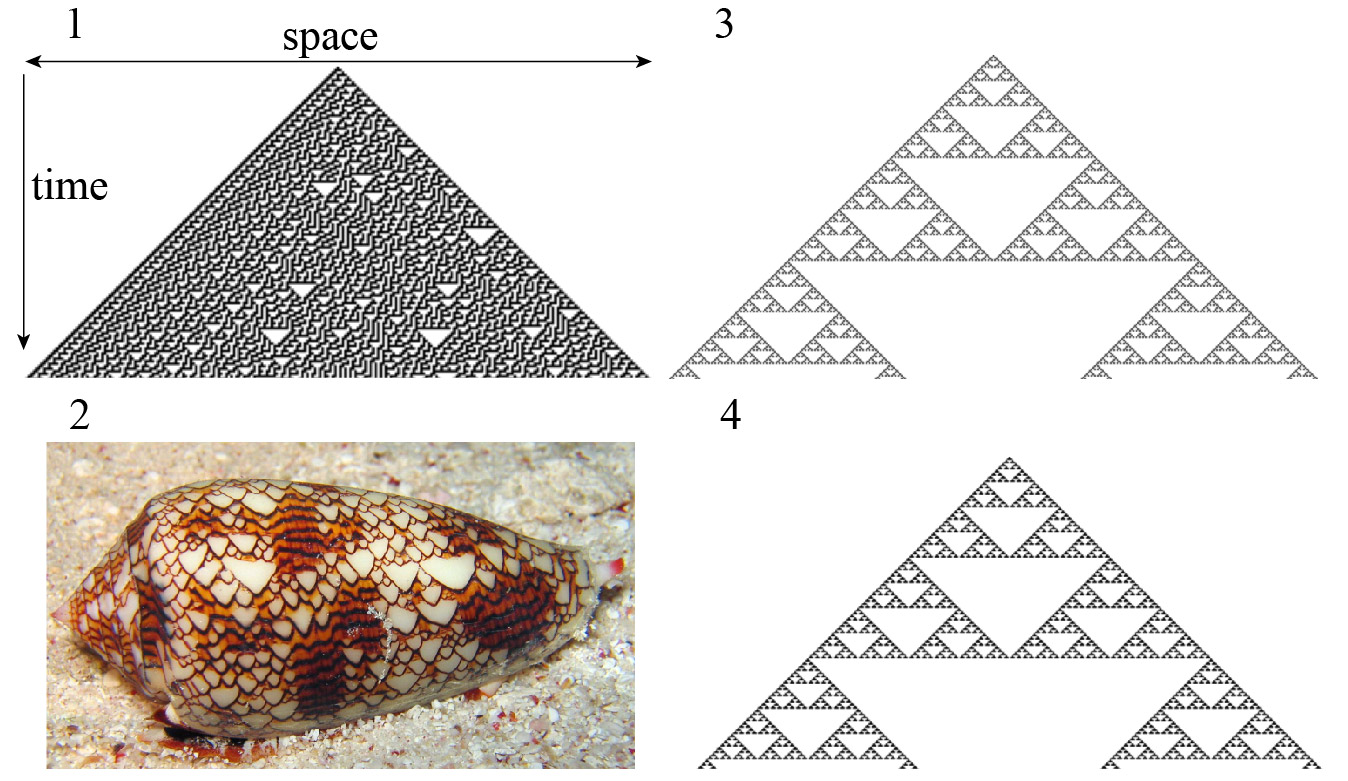}
  \caption{(1) and (2): Elementary 1D CA Rule 30 exhibits a pattern resembling the natural shell of \textit{Conus textile}; (3) and (4): Elementary 1D CA rule 30 and 2D Moore neighborhood outer totalistic CA rule 510 both exhibit Sierpiński triangle-like patterns. These CA patterns are visualized as space-time diagrams. (Image (2) adapted from ~\cite{shell})}
  \label{fig_someCAs}
\end{figure}

Wolfram’s CA numbering system \cite{Wolfram_new} categorizes all possible two-state CA configurations in both 1D and 2D \cite{Wolfram_2D}. In 1D CA, each cell updates its state based on its current state and those of its left and right neighbors. Figure~\ref{fig_someCAs} (1) and (2) show Rule 30, which displays aperiodic, chaotic behavior, visually similar to the natural shell of \textit{Conus textile} \cite{shell}. In contrast, Rule 90 exhibits a self-similar pattern known as the Sierpiński triangle (Figure~\ref{fig_someCAs} (3)).

Most discussions of self-similarity in CA focus on 1D space-time diagrams, where time progresses vertically and recursive patterns emerge through generations. While certain 2D CA, such as the one shown in Figure~\ref{fig_someCAs} (4), also produce similar fractal patterns, demonstrating that Sierpiński-like structures can appear in cross-sections of 2D grids evolving over time.

As demonstrated in Wolfram and co-workers’ systematic studies, such self-similar features are rarely observed in purely spatial diagrams. This makes the UWCA particularly unique, as it reveals recursive geometric structures that suggest a form of spatial self-similarity, not merely arising from temporal evolution.
%In 2D CA, the grid expands to two dimensions, with each cell surrounded by a neighborhood---commonly the Moore neighborhood (eight surrounding cells) or von Neumann neighborhood (four orthogonally adjacent cells).

%For example, In Conway's Game of Life, each cell's next state is determined by the number of live neighbors in Moore neighborhood, following simple rules that lead to emergent behaviors like oscillators, gliders, and self-replicating structures. 

%***[purpose unclear]Both 1D and 2D CA demonstrate how local interactions governed by simple rules can give rise to global complexity, forming the conceptual foundation for various visual systems complex systems research. They found out a number of rules exhibit self-similar fractal patterns on space-time diagrams, with only one cell live in the initial configuration. Such as Rule 90 in elementary 1D CA (Figure~\ref{fig_someCAs} (3)), and 2D Moore neighborhood outer totalistic CA rule 510 (Figure~\ref{fig_someCAs} (4)). 

\subsection{Ulam-Warburton Cellular Automaton}
\begin{figure}[h]
  \centering
  \includegraphics[width=\linewidth]{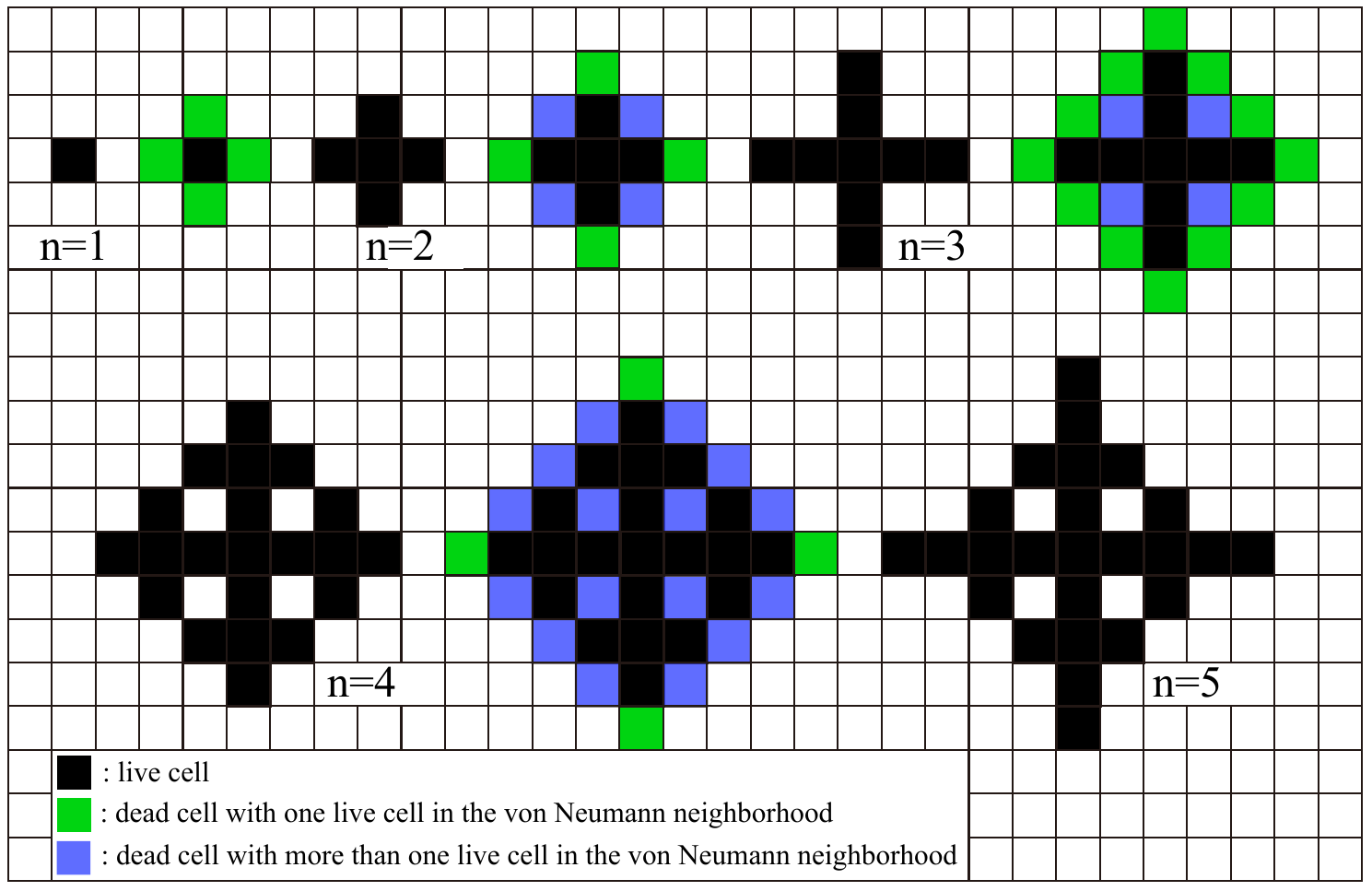}
  \caption{Visualization of UWCA from generations $n = 1$ to $5$. The initial state ($n = 1$) begins with a single live cell. Within each generation, dead cells evaluate their von Neumann neighborhood (up, down, left, right). If exactly one neighbor is alive (green), the cell becomes alive in the next generation. Cells with more than one live neighbors (blue) indicate invalid transitions.}
  \label{fig_UWCA}
\end{figure}

Ulam-Warburton Cellular Automaton (UWCA) can also be described using Wolfram code as 5-neighbor outer totalistic rule, represented by codes 686. Its evolution can be explained in few informal steps, as visually explain in Figure~\ref{fig_UWCA}.

\begin{enumerate}
    \item Create a two-dimensional grid with all cells in a dead state (white).
    \item For the first generation, initialize a single live cell (black).
    \item For each dead cell, if exactly one of its von Neumann neighborhood cell is alive, it becomes alive in the next generation.
    \item Iterate the generation, and repeat from step (3).
\end{enumerate}

UWCA also displays self-similarity pattern in its space-time diagram, similar to rule 150 in elementary CA \cite{Kawaharada_rule150}. However, whether it exhibits a self-similar spatial pattern (purely from a 2D top-down view) remains a subject of ongoing debate.

Applegate and co-workers highlighted UWCA's recursive features both geometrically and arithmetically \cite{Applegate_tooth}. For instance, they observed that the pattern of live (black) cells formed a square every $2^k$ generations ($k \in \mathbb{N}$), and that the number of newly activated live cells per generation could be described by a recursive function, where $u(n)$ denoted the number of cells that changed from dead to live at the $n$th generation.
% Requires: \usepackage{amsmath}

\begin{equation}
\label{eq:UWCA}
\begin{aligned}
    & u(0) = 1,\quad u(1) = 1,\quad \text{for } k \ge 0, \\
    & u(2^{k} + 1 + i) =
    \begin{cases}
        4, & \text{if } i = 0, \\
        3u(i+1), & \text{if } i = 1, \ldots, 2^{k}-1.
    \end{cases}
\end{aligned}
\end{equation}

\begin{figure}[h]
  \centering
  \includegraphics[width=\linewidth]{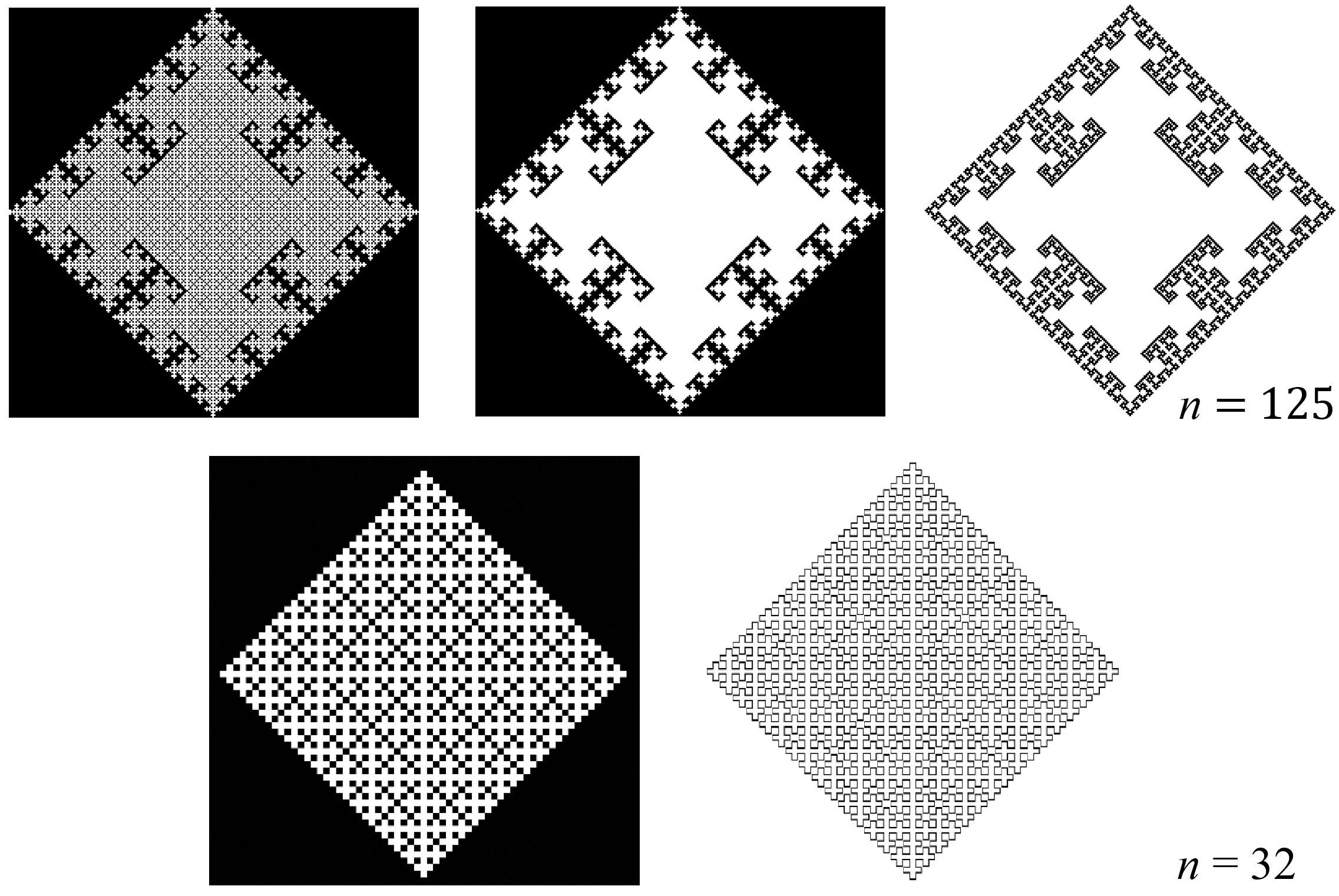}
  \caption{Top-left: UWCA at generation $n=125$; Top-middle: Using image processing to extract only the boundary contours of the pattern; Top-right: Edge feature extracted with edge detection method. Bottom-left: At generation $n=32$. Bottom-right: Edge feature extracted with edge detection method \cite{Kawaharada}.}
  \Description{UWCA pattern in Kawaharada's research.}
  \label{fig_Kawa}
\end{figure}

They argued that the UWCA exhibited only fractal-like patterns in its spatial diagrams. In contrast, Kawaharada provided evidence that UWCA did form true self-similar fractals under certain conditions \cite{Kawaharada}. As shown in Figure~\ref{fig_Kawa}, they demonstrated this using the box-counting method \cite{BoxCounting}, showing that at generation $2^5=32$, the structure formed a perfect square with a fractal dimension of 2, while at generation $2^7-3=125$, the measured fractal dimension was approximately 1.585. This finding suggested that UWCA exhibited conditional fractal behavior in spatial form, depending on the generation we choose. Overall, these findings indicate that UWCA's spatial fractal characteristics do not accumulate consistently across generations, but instead represent under specific generations.

\section{Recursive Gradient Profiling Procedure}
Although prior studies have shown that the UWCA exhibits recursive growth and conditional fractal behavior at specific generations, its self-similar structure remains difficult to perceive in purely spatial, binary renderings. By collapsing temporal information into uniform states, conventional visualizations obscure how recursion accumulates over time. To address this, we introduce a visualization-driven design method that re-encodes generational information as spatial intensity, revealing latent self-similarity cumulatively.

Applegate and co-workers~\cite{Applegate_tooth} analyzed the geometric and arithmetic properties of UWCA, revealing the recursive behavior of its live cell configurations. Specifically, they showed that every $2^k$ generation (for $k \in \mathbb{N}$), the set of live cells forms a rectangular pattern, and the number of newly activated cells at each generation can be described by Equation ~\ref{eq:UWCA}. These findings suggest an underlying order in UWCA's growth that aligns with patterns found in fractal structures.

Building on this observation, we propose a visualization method that enhances the visibility of such recursive patterns. In particular, We treat each range between $2^k$ and $2^{k+1}$ generations as a grouped sequence to visualize the growth trend within each group using a continuous gradient profile. Instead of assigning a constant black or white value, we normalize each generation’s newly born live cells to a grayscale intensity between 0 and 1, allowing for smooth transitions that reveal recursive growth patterns over time.

To achieve this, we define a Recursive Gradient Profile Function (RGPF), which assigns grayscale values to each generation in a given group, as formalized in Equation ~\eqref{eq:grayscale}:

\begin{equation}
\text{grayscale intensity } f(n) =
\begin{cases}
0, & n < 1 \\
2 - n, & 1 \leq n \leq 2 \\
f\left( \dfrac{n}{2} \right), & n > 2
\end{cases}
    \label{eq:grayscale}
\end{equation}

\begin{figure}[h]
  \centering
  \includegraphics[width=\linewidth]{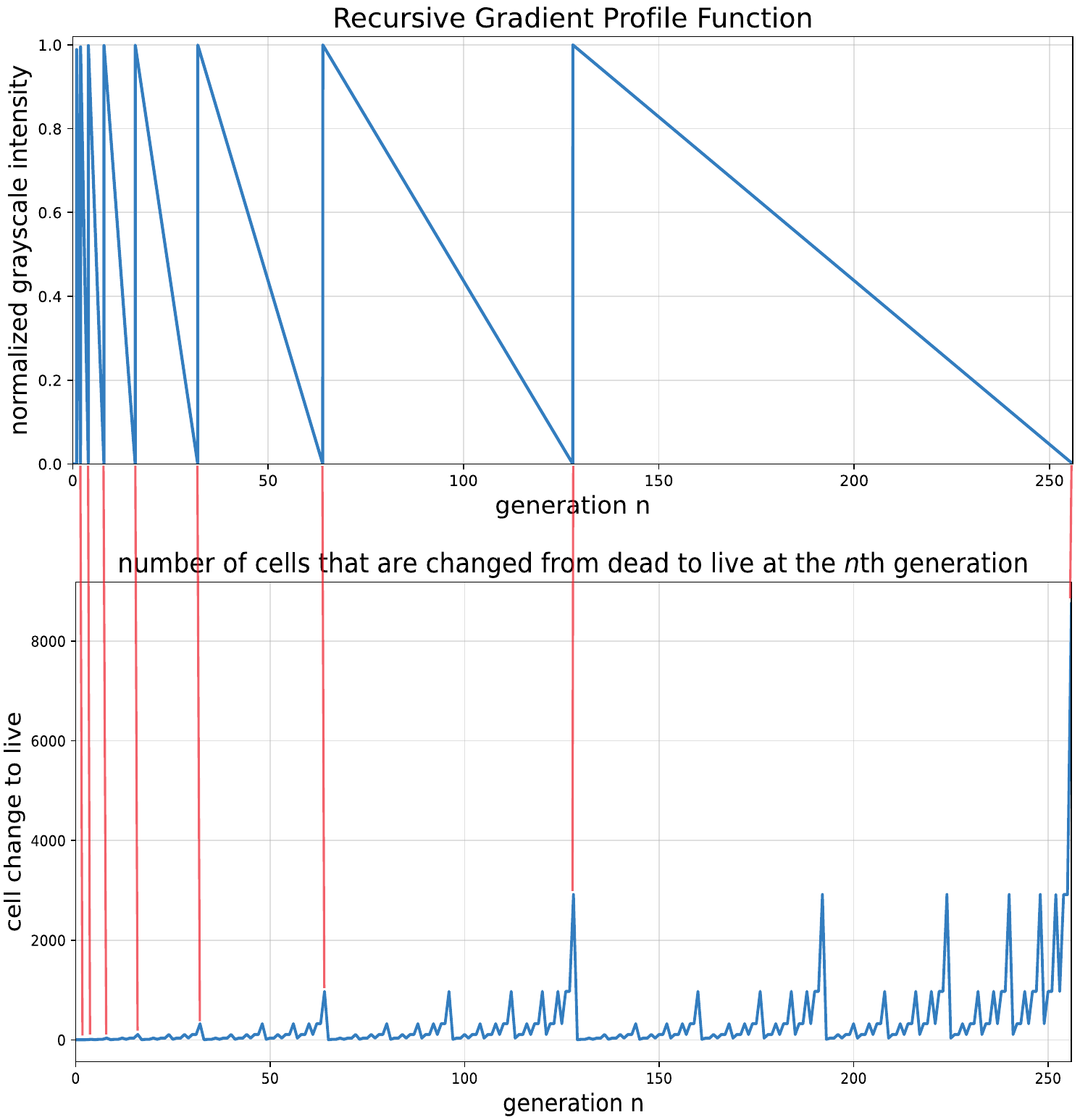}
  \caption{Top: Mapping of generation index ($n=1$ to $256$) to grayscale intensity values based on the Recursive Gradient Profile Function (RGPF), as defined in Equation~\eqref{eq:grayscale}. Bottom: Number of newly born live cells at each generation, computed using the UWCA update rule in Equation~\eqref{eq:UWCA}. Red lines indicate when $n=2^k$. At these generations, live cells form square patterns.}
  \Description{Plot of the recursive function}
  \label{fig_uwca_rgpf}
\end{figure}

\begin{figure}
  \centering
  \includegraphics[width=\linewidth]{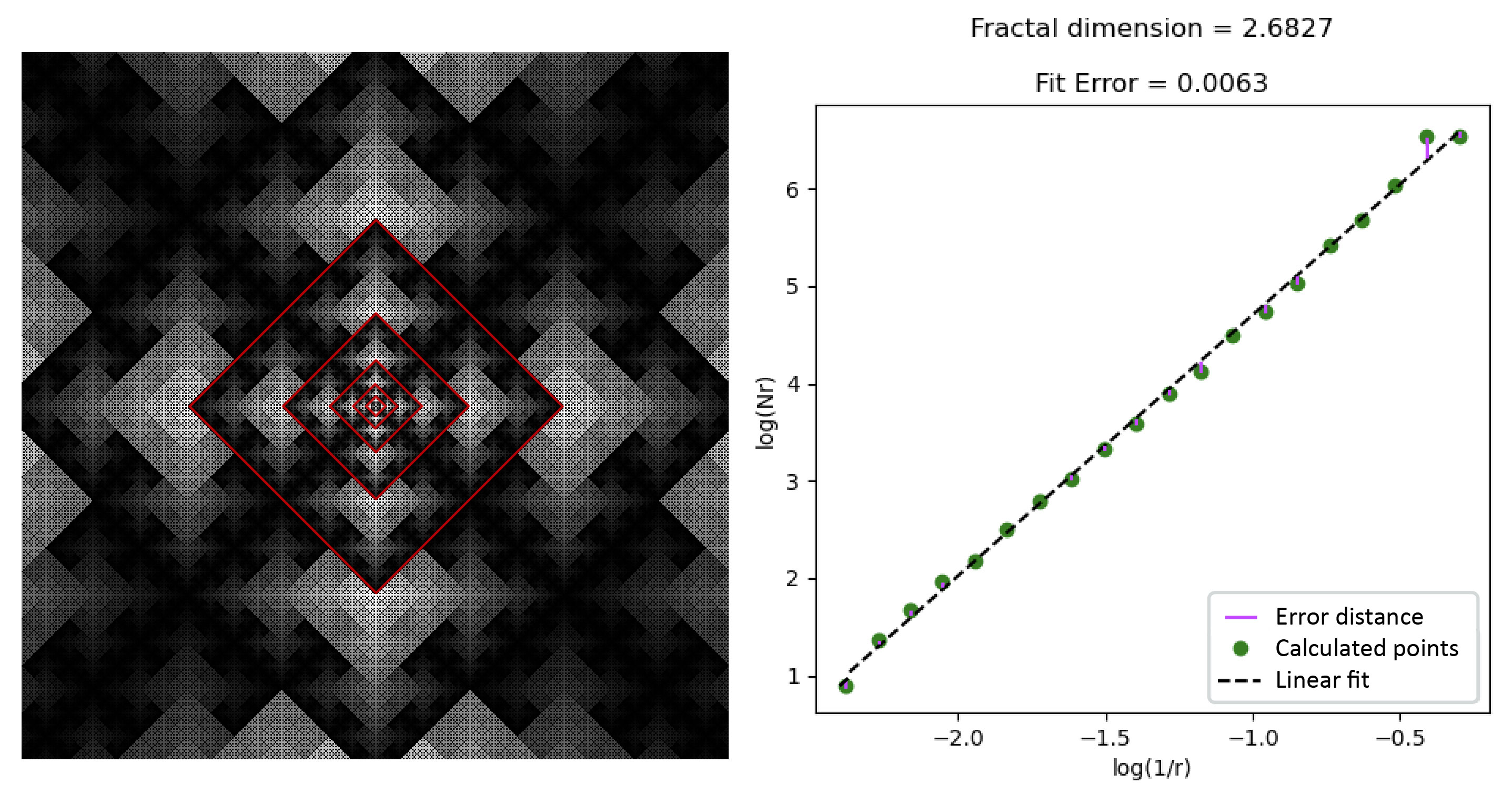}
  \caption{
Left: Shows the UWCA multiplying recursive gradient profile, and output up to generation 256; The red squares refer to Figure~\ref{fig_uwca_rgpf} red lines, indicating generations where $n = 2^k$ and square outlines emerge. 
Right: Displays the corresponding log–log plot, from which a fractal dimension of $D = 2.6827$ is computed. 
Fit error $E = 0.0063$ quantifies the average vertical distance between the data points and the regression line. 
The error distance lines representing the vertical distance from each point to the fitted line, visualizing the fitting quality.
}

  \Description{Plot of the recursive function}
  \label{fig_sdbc}
\end{figure}

The top figure in Figure~\ref{fig_uwca_rgpf} shows RGPF for $n=1$ to $256$, while the bottom figure shows the total number of cells that changed from dead to live at generation $n$ based on Equation~\eqref{eq:UWCA}. The red lines shows how the sharp edges of the intensity profile relates to the generation of new live cells in the pattern. As one can see, the sharp rises of the gradient profile corresponds to spawns of new cell generation. These moments are indicated by red squares in Figure~\ref{fig_sdbc} (left).

Figure~\ref{fig_sdbc} (left) shows the result of multiplying recursive gradient profile to all newly activated live cells up to the 256th generation. Each cell corresponds to one pixel, resulting in an output resolution of 480~$\times$~480 pixels. We analyzed the fractal dimension of the pattern using the Shift Differential Box Counting (SDBC) method~\cite{Chen_SDBC}. Compared to the standard Differential Box Counting (DBC) approach, SDBC improves accuracy through two key enhancements. First, it introduces a computation-quantization process that vertically shifts boxes along the gray-level (z) axis, minimizing overcounting and undercounting errors. Second, it is gray-level shift invariant, meaning its estimated fractal dimension remains stable under changes in brightness, making it robust for analyzing grayscale textures and gradient-based structures such as UWCA with gradient profiling.

Figure~\ref{fig_sdbc} (right) shows the log–log plot to calculate the fractal dimension, resulting in $D = 2.6827$ with a fitting error of $E = 0.0063$. In this plot, the horizontal axis represents $\log(1/r)$, where $r$ is the scaling ratio $r=s/M$, with $s$ being the size of each box and $M$ the total image size. As $r$ decreases, $1/r$ reflects the level of spatial subdivision. The vertical axis shows $\log(N_r)$, where $N_r$ is the number of boxes required to cover the grayscale surface at each scale. The slope of the regression line in this log–log space corresponds to the fractal dimension $D$, capturing how detail in the image of different scales.

The error $E$ corresponds to the root mean square (RMS), which measures the average vertical distance between data points and the regression line in log-log space. This value reflects how well the data follows a linear trend. In order to measure its quality, we then normalized $E$:
\begin{equation}
E_{\text{norm}} = \frac{E}{\log N_r^{\max} - \log N_r^{\min}}
\end{equation}
where $\log N_r^{\max}$ and $\log N_r^{\min}$ are 6.532 and 0.903, respectively. The resulting normalized error is $E_{\text{norm}} \approx 0.112\%$. Such a low value indicates an excellent fit, confirming the consistent of self-similarity across scales.

% The error $E$ corresponds to the root mean square (RMS) distance between the data points and the regression line, it refers as the average vertical distance from each data point to the regression line in log–log space. This value reflects how well the data follows a linear trend—a key property of fractals in logarithmic scale. An $E$ value as low as 0.0063 indicates an excellent fit, confirming the presence of consistent self-similar structure across scales. *** [convert to percentage] ** 

\begin{figure}[h]
  \centering
  \includegraphics[width=\linewidth]{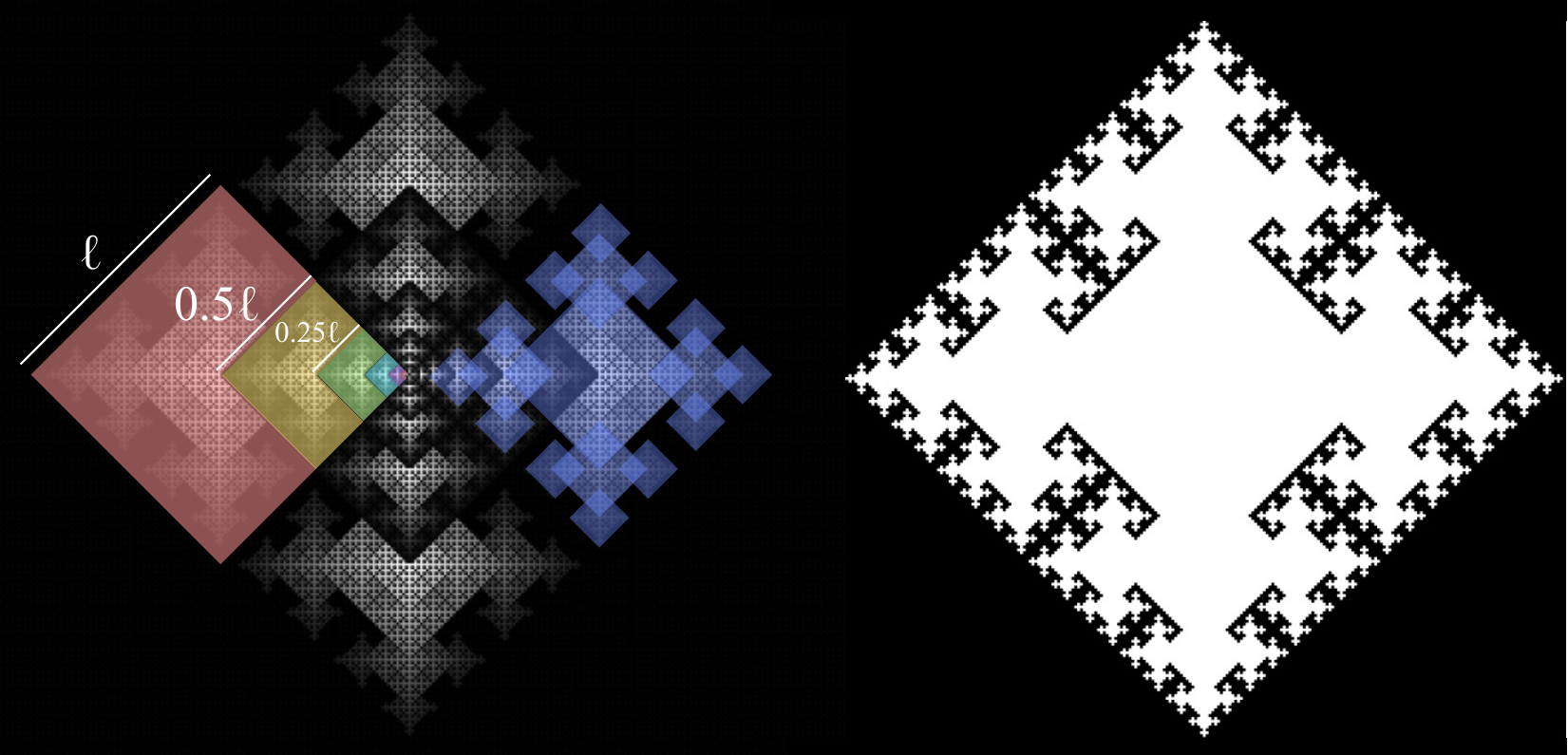}
  \caption{Comparison between the gradient-adjusted pattern (left) and the original UWCA visualization by Kawaharada \cite{Kawaharada} (right). In the gradient-adjusted pattern, the left masks highlight self-similar structures formed by recursively scaled squares, while the right blue mask reveals overlapping square clusters resembling a crystal-like recursive pattern. These features are not apparent in uniform-color renderings but become visible through recursive gradient profiling.}
  \Description{Plot of the recursive function}
  \label{fig_uwca_Kawa}
\end{figure}

In order to compare our generated results with the visualizations by Kawaharada \cite{Kawaharada}, we highlight the self-similar fractal pattern that were not visible in the original UWCA visualizations, where all live cells were rendered in a uniform color). Figure~\ref{fig_uwca_Kawa} (left) shows a gradient-adjusted UWCA overlaid with square masks of doubling sizes, revealing the self-similar characteristic of the pattern. On the right, a blue mask highlights clusters of squares of varying sizes, forming a crystal-like recursive pattern. These visual patterns are easily visible through recursive gradient profiling and cannot be observed through the original diagrams, as shown in ~\ref{fig_uwca_Kawa} (right).

%As shown in Figure~\ref{fig_uwca&Kawa} (left), on the UWCA applying RGPF, a semi-transparent mask illustrates a set of squares that recursively form larger squares through a two-fold scaling relationship with their neighbors, which refers to the fundamental characteristic of self-similar pattern. 

Building on the initial UWCA design, we further examine it under alternative neighborhood conditions. Previous studies of UWCA have focused on two classical neighborhood configurations: the von Neumann and Moore neighborhoods, which are the most common in CA's neighborhood design. In the broader CA literature, various neighborhood models have been explored, including displaced von Neumann, Smith, and Cole neighborhoods~\cite{Batty_CA, cole, circular_CA}.

\begin{figure*}[h]
  \centering
  \includegraphics[width=0.95\linewidth]{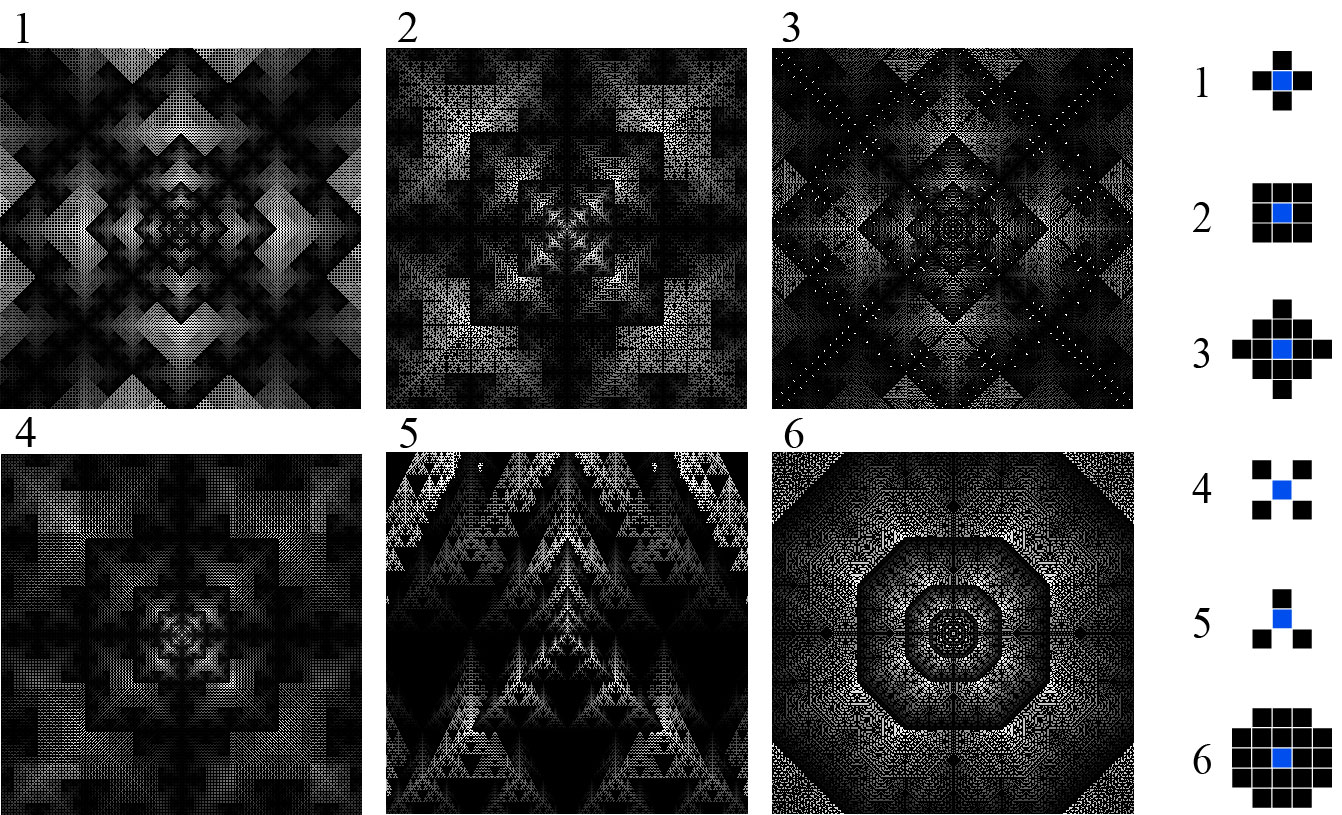}
  \caption{Evolution of different UWCA neighborhood configurations up to $n=256$, including: 1. von Neumann, 2. Moore, 3. Moore von Neumann, 4. displaced von Neumann, 5. Cole, 6. Circular.}
  \Description{Evolution of different UWCA neighborhood configurations up to $n=256$, including: 1. von Neumann, 2. Moore, 3. Moore von Neumann, 4. displaced von Neumann, 5. Cole, 6. Circular.}
  \label{fig_manyUWCA}
\end{figure*}

We extend the UWCA framework by exploring several alternative neighborhood configurations while preserving both the original UWCA rule and the gradient profiling. Figure~\ref{fig_manyUWCA} illustrates that multiple neighborhood variants generate visually distinct yet structurally comparable patterns. We further quantify their geometric properties by measuring fractal dimensions and fitting errors using the SDBC method. As summarized in Table~\ref{tab:neighborhood_de}, all six neighborhood variants yield fractal dimensions $D$ consistently between 2 and 3, in accordance with the definition of fractal geometry. Moreover, the normalized fitting errors remain low across all configurations (with all $E_{\text{norm}} \leq 0.12\%$), indicating strong linear behavior in log-log space. These results demonstrate our gradient profile reveals self-similar fractal structures across different neighborhoods, underscoring both the robustness and the visual richness of the extended UWCA framework.

\begin{table}[ht]
    \centering
    \caption{Neighborhood configurations with corresponding $D$, $E$ and $E_{\text{norm}}$ values}
    \label{tab:neighborhood_de}
    \begin{tabular}{lccc}
        \toprule
        Neighborhood & $D$ & $E$ & $E_{\text{norm}}$ (\%) \\
        \midrule
        1. Von Neumann & 2.6827 & 0.0063 & 0.112\\
        2. Moore & 2.7072 & 0.0066 & 0.116 \\
        3. Moore von Neumann & 2.6519 & 0.0150 & 0.116 \\
        4. Displaced von Neumann & 2.7150 & 0.0069 & 0.12 \\
        5. Cole & 2.6499 & 0.0062 & 0.111 \\
        6. Circular & 2.7492 & 0.0058 & 0.102  \\
        \bottomrule
    \end{tabular}
\end{table}

\section{Computational Systems to Cultural Forms: Recursive Gradient Profiling as a Bridge}
%\section{Discussion **** name change***}
Beyond exploring the relationship between CA and fractal geometry, this study also seeks a potential intersection between computational systems and cultural art. In Figure~\ref{fig_uwca_Kawa}, the UWCA demonstrates a recursive spatial growth pattern. Starting from a single seed cell, its structure expands outward while continually reproducing its original shape at larger scales. If the evolution continues indefinitely, one can imagine it as exhibiting self-similarity at arbitrary scales, a principle reminiscent of the infinite mirror and video feedback~\cite{fractal_book_Peitgen}, where repeated reflections at varying scales maintain a consistent structural resemblance.

\begin{figure*}
  \centering
  \includegraphics[width=0.95\linewidth]{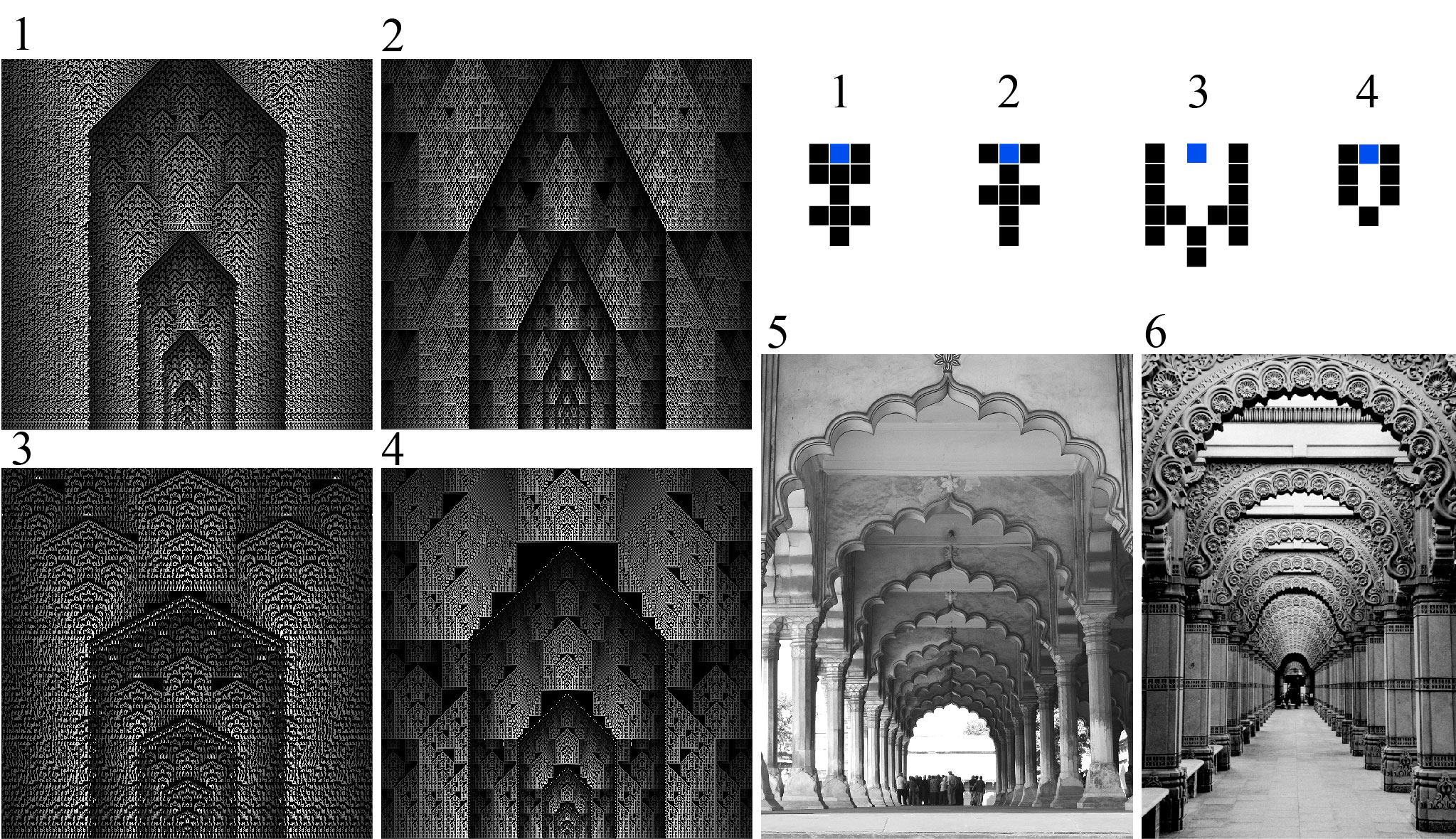}
  \caption{(1)\textasciitilde(4): UWCA generated with recursive gradient profiling for various neighborhoods. (5): Agra Fort Agra (\textcopyright~Herojit Waikhom 2012, via Wikimedia Commons) and (6): BAPS Swaminarayan Akshardham continus arch structures (\textcopyright~Vivek Varma 2023, via Unsplash).
The UWCA pattern resembles the sequential arrangement of arches in Arabic and Hindu temples. In both images, the arches scale proportionally as they recede into space, eventually converging at a vanishing point. This visual rhythm closely aligns with the self-similarity property inherent in the UWCA structure.}
  \Description{Plot of the recursive function}
  \label{fig_archUWCA}
\end{figure*}

\begin{figure*}
  \centering
  \includegraphics[width=0.75\linewidth]{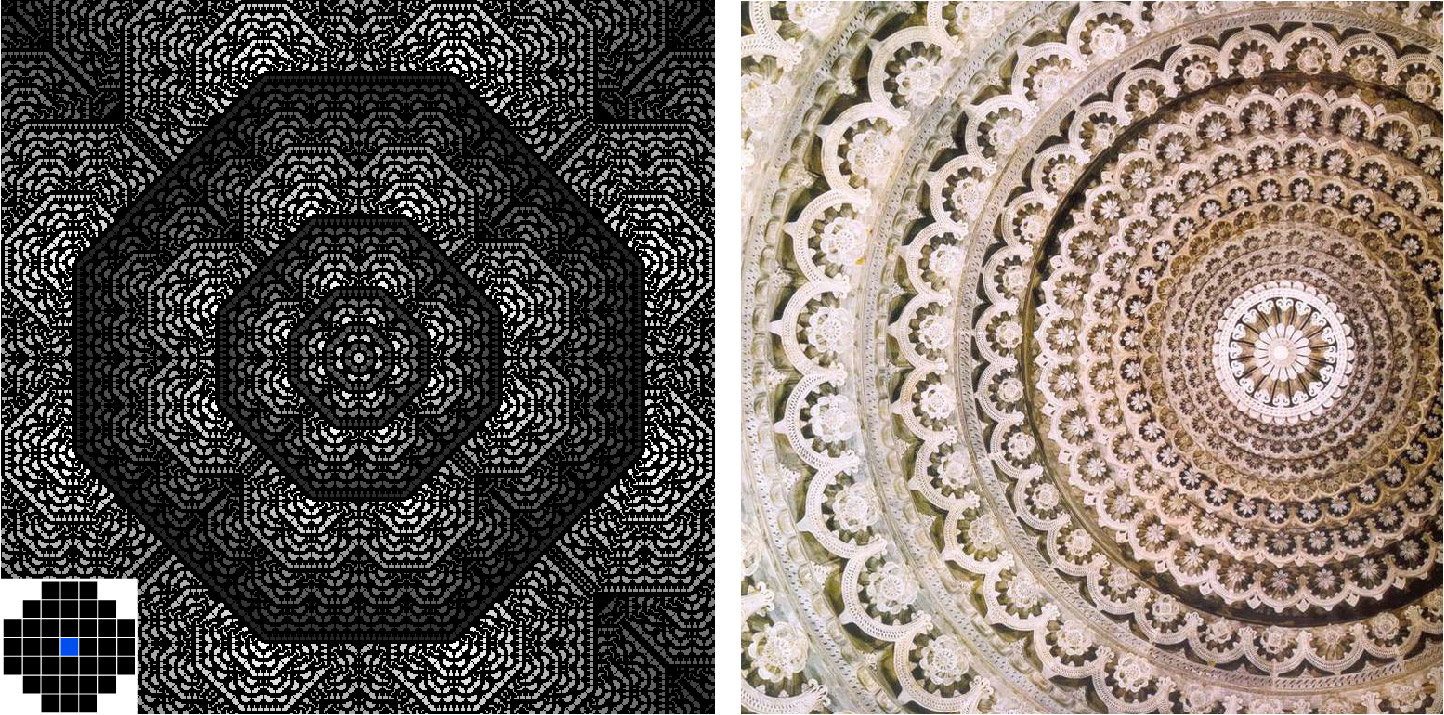}
  \caption{Left: UWCA generated using an alternative neighborhood with gradient profiling. Right: Ceiling dome of the Lila Mandapam~\cite{lila}. The dome's concentric patterns expand from the center, forming a radial visual structure. Similarly, the UWCA pattern on the left reveals a fan-like shape emerging from the central point, created through recursive growth resonating with the dome geometric expansion.}
  \Description{Plot of the recursive function}
  \label{domeUWCA}
\end{figure*}

This concept also parallels certain architectural visual effects, as seen in the Figure~\ref{fig_archUWCA} (5) and (6), where repeating forms create a perceived infinite recession toward a vanishing point. Such recursive spatial rhythms evoke a sense of visual infinity. Our recursive gradient profiling reveals similar self-similarity in UWCA, making these geometric recursions more perceptible and resonant with motifs found in religious architecture. In particular, India's Swaminarayan Akshardham temple complex, an UNESCO World Heritage Site, features multiple mandapam halls, each crowned with intricately ornamented domes that exhibit a fractal-like aesthetic. Figure~\ref{domeUWCA} shows the dome of the Lila Mandapam as an example. Another case is the Khajuraho Group of Monuments, also a UNESCO World Heritage Site, known for its Hindu and Jain temples. These structures are distinguished by richly carved surfaces and repetitive arrangements across walls, domes (mandapas), and towers (shikharas), producing effects that can be interpreted as fractal patterns in both plan and elevation. These patterns not only serve aesthetic purposes but also reflect cosmological and mathematical symbolism rooted in Hindu temple design.

\begin{figure*}
  \centering
  \includegraphics[width=0.9\linewidth]{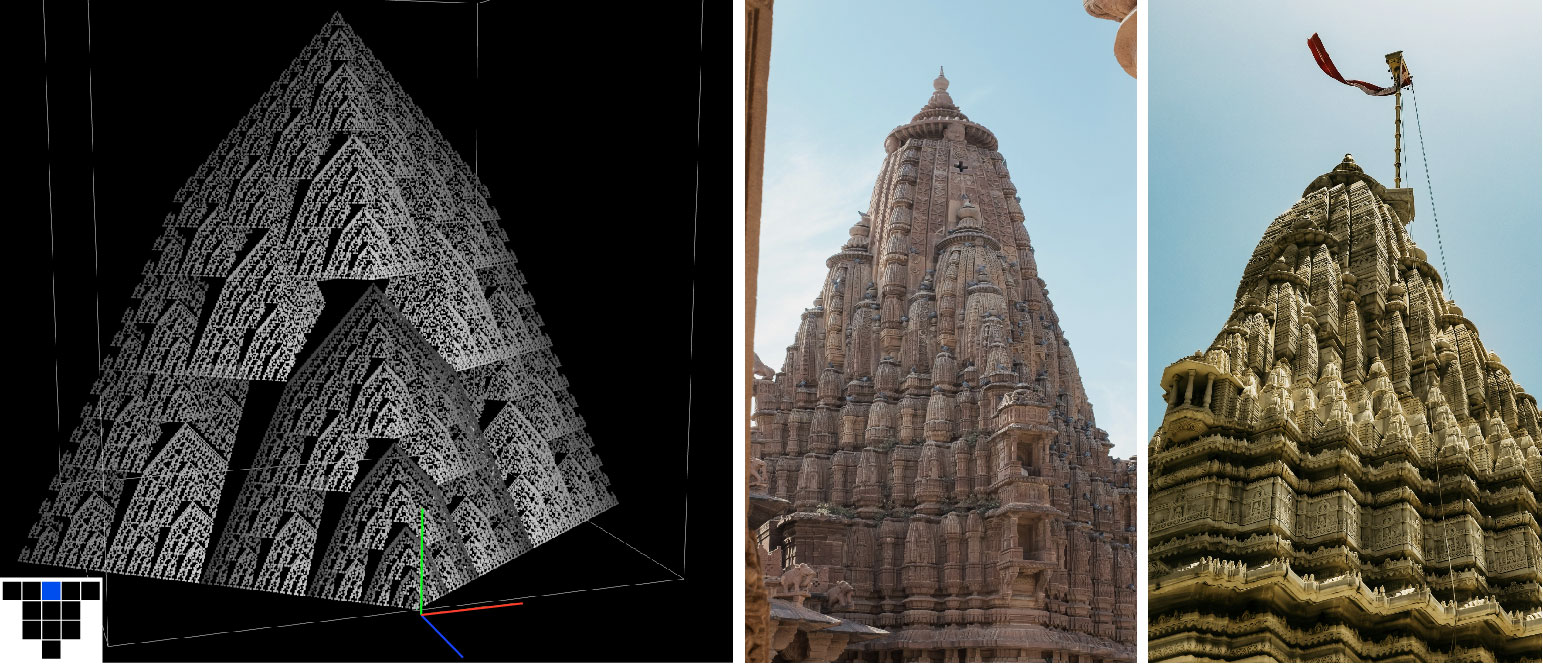}
  \caption{Left: Space-time visualization of UWCA using alternative neighborhood, with each generation growing in the negative direction of blue axis). Middle and right: Hindu temples of the Adani Shantigram Jain Temple (\textcopyright~Laura Lezman 2025, via Unsplash) and Mandore (\textcopyright~Yudhajit Ghosh 2025, via Unsplash). 
By conceptualizing time as a vertical axis in the space-time diagram, each layer of newly born cells forms a tiered, stepwise structure. This visual stacking resembles the recursive elevation seen in temples. Both share modular repetition across scales, suggesting a visual analogy between algorithmic recursion and symbolic architectural layering.}
  \Description{Plot of the recursive function}
  \label{templeUWCA}
\end{figure*}

Figure~\ref{templeUWCA} (left) shows a space-time visualization of UWCA using alternative neighborhoods, where newly activated cells grow generation by generation toward the negative direction of blue axis. Beside it, we juxtapose images of selected Hindu temples, including Adani Shantigram Jain Temple (middle) and Mandore Temple (right), inviting a visual dialogue between algorithmic recursion and sacred architectural geometry.

Through these comparative visual examples, we reveal new expressive potentials for CA systems, not only in their applications of simulating natural phenomena, but also in generating cultural, symbolic, and aesthetically rich fractal patterns that bridge complexity science and cultural art.

\section{Conclusion}

Originally conceived by von Neumann and Ulam to model natural growth via local rules, CA have become a foundational tool in complex systems research, with applications across computer science, physics, earth sciences, and generative art.

Ulam-Warburton Cellular Automaton (UWCA), one of the earliest CA models, influenced tree growth simulations in computer graphics. This study revisits UWCA from a visual perspective, focusing on its recursive spatial expansion. Since self-similar fractal patterns in cellular automata typically emerge within space-time diagrams rather than in purely spatial dimensions, we investigate alternative visual strategies to reveal fractal characteristics often hidden in traditional renderings.

By grouping cell generations between $2^k$ and $2^{k+1}$, we introduced a Recursive Gradient Profile Function (RGPF) to apply linear gradient shading across generations. While traditional black-and-white visualizations of UWCA only reveal self-similar fractal structures at boundaries and specific generations, this gradient profile enables a continuous and cumulative expression of self-similarity embedded within the recursive growth.

We further extended the UWCA by applying alternative neighborhood configurations beyond original design, including Displaced von Neumann, Cole, and Circular neighborhoods. When combined with our gradient profile, these variants continued to produce self-similar fractal patterns. We evaluated the fractal dimensions of their grayscale outputs using the Shift Differential Box Counting (SDBC) method. All configurations yielded fractal dimensions $D$ between 2 and 3---falling between the topological dimensions of a surface and a volume---and maintained low fitting errors ($E_{\text{norm}} \leq 0.12\%$), indicating strong fractal characteristics across the variants.

These recursive patterns resonate with geometric repetition, optical effects like infinite mirrors and video feedback, and recursive aesthetics found in both historical and cultural contexts. They echo the concept of \textit{mise en abyme} in Western art history and align with fractal motifs observed in religious architectural forms.

Beyond UWCA, the recursive gradient visualization approach holds potential for other CA models with similar growth rules, such as series of toothpick sequence CA. Overall, this work suggests new possibilities for connecting complex systems science with cultural art and design, offering a generative framework that bridges computation, visual design, and artistic interpretation.


% \begin{acks}
% \end{acks}

\bibliographystyle{ACM-Reference-Format}
\bibliography{references}


\begin{thebibliography}{31}

%%% the Web address.

\ifx \showCODEN    \undefined \def \showCODEN     #1{\unskip}     \fi
\ifx \showISBNx    \undefined \def \showISBNx     #1{\unskip}     \fi
\ifx \showISBNxiii \undefined \def \showISBNxiii  #1{\unskip}     \fi
\ifx \showISSN     \undefined \def \showISSN      #1{\unskip}     \fi
\ifx \showLCCN     \undefined \def \showLCCN      #1{\unskip}     \fi
\ifx \shownote     \undefined \def \shownote      #1{#1}          \fi
\ifx \showarticletitle \undefined \def \showarticletitle #1{#1}   \fi
\ifx \showURL      \undefined \def \showURL       {\relax}        \fi
% invisible to TeX
\providecommand\bibfield[2]{#2}
\providecommand\bibinfo[2]{#2}
\providecommand\natexlab[1]{#1}
\providecommand\showeprint[2][]{arXiv:#2}

\bibitem[Acosta(2025)]%
        {Acosta_recursive}
\bibfield{author}{\bibinfo{person}{Gabriel Acosta}.} \bibinfo{year}{2025}\natexlab{}.
\newblock \bibinfo{booktitle}{\emph{The Recursive Architecture of Reality: How Love Operates the Universe}}.
\newblock Principia Recursionis.
\newblock
\urldef\tempurl%
\url{https://www.principiarecursionis.com/post/the-recursive-architecture-of-reality-how-love-operates-the-universe}
\showURL{%
\tempurl}
\newblock
\shownote{Accessed: 2025-12-29}.


\bibitem[Applegate et~al\mbox{.}(2010)]%
        {Applegate_tooth}
\bibfield{author}{\bibinfo{person}{David Applegate}, \bibinfo{person}{Omar~E Pol}, {and} \bibinfo{person}{Neil~JA Sloane}.} \bibinfo{year}{2010}\natexlab{}.
\newblock \showarticletitle{The toothpick sequence and other sequences from cellular automata}.
\newblock \bibinfo{journal}{\emph{arXiv preprint arXiv:1004.3036}} (\bibinfo{year}{2010}).
\newblock


\bibitem[Batty(2000)]%
        {Batty_geo}
\bibfield{author}{\bibinfo{person}{Michael Batty}.} \bibinfo{year}{2000}\natexlab{}.
\newblock \showarticletitle{Geocomputation using cellular automata}. In \bibinfo{booktitle}{\emph{Geocomputation}}. New York: Taylor \& Francis, \bibinfo{pages}{95--126}.
\newblock


\bibitem[Batty(2007)]%
        {Batty_CA}
\bibfield{author}{\bibinfo{person}{Michael Batty}.} \bibinfo{year}{2007}\natexlab{}.
\newblock \bibinfo{booktitle}{\emph{Cities and complexity: understanding cities with cellular automata, agent-based models, and fractals}}.
\newblock \bibinfo{publisher}{The MIT press}.
\newblock


\bibitem[Bunde and Havlin(2013)]%
        {fractal_book_Bunde}
\bibfield{author}{\bibinfo{person}{Armin Bunde} {and} \bibinfo{person}{Shlomo Havlin}.} \bibinfo{year}{2013}\natexlab{}.
\newblock \bibinfo{booktitle}{\emph{Fractals in science}}.
\newblock \bibinfo{publisher}{Springer}.
\newblock


\bibitem[Cannon(1984)]%
        {Cannon_fractal}
\bibfield{author}{\bibinfo{person}{JW Cannon}.} \bibinfo{year}{1984}\natexlab{}.
\newblock \showarticletitle{The fractal geometry of nature. by Benoit B. Mandelbrot}.
\newblock \bibinfo{journal}{\emph{The American Mathematical Monthly}} \bibinfo{volume}{91}, \bibinfo{number}{9} (\bibinfo{year}{1984}), \bibinfo{pages}{594--598}.
\newblock


\bibitem[Chen et~al\mbox{.}(2003)]%
        {Chen_SDBC}
\bibfield{author}{\bibinfo{person}{Wen-Shiung Chen}, \bibinfo{person}{Shang-Yuan Yuan}, {and} \bibinfo{person}{Chih-Ming Hsieh}.} \bibinfo{year}{2003}\natexlab{}.
\newblock \showarticletitle{Two algorithms to estimate fractal dimension of gray-level images}.
\newblock \bibinfo{journal}{\emph{Optical Engineering}} \bibinfo{volume}{42}, \bibinfo{number}{8} (\bibinfo{year}{2003}), \bibinfo{pages}{2452--2464}.
\newblock


\bibitem[Conway et~al\mbox{.}(1970)]%
        {Conway_game}
\bibfield{author}{\bibinfo{person}{John Conway} {et~al\mbox{.}}} \bibinfo{year}{1970}\natexlab{}.
\newblock \showarticletitle{The game of life}.
\newblock \bibinfo{journal}{\emph{Scientific American}} \bibinfo{volume}{223}, \bibinfo{number}{4} (\bibinfo{year}{1970}), \bibinfo{pages}{4}.
\newblock


\bibitem[Coombes(2009)]%
        {shell}
\bibfield{author}{\bibinfo{person}{Stephen Coombes}.} \bibinfo{year}{2009}\natexlab{}.
\newblock \showarticletitle{The geometry and pigmentation of seashells}.
\newblock \bibinfo{journal}{\emph{Nottingham: Department of Mathematical Sciences, University of Nottingham}} (\bibinfo{year}{2009}).
\newblock


\bibitem[Falconer(2013)]%
        {fractal_book_Falconer}
\bibfield{author}{\bibinfo{person}{Kenneth Falconer}.} \bibinfo{year}{2013}\natexlab{}.
\newblock \bibinfo{booktitle}{\emph{Fractal geometry: mathematical foundations and applications}}.
\newblock \bibinfo{publisher}{John Wiley \& Sons}.
\newblock


\bibitem[Foroutan-pour et~al\mbox{.}(1999)]%
        {Firiytab_fractalDimension}
\bibfield{author}{\bibinfo{person}{Kayhan Foroutan-pour}, \bibinfo{person}{Pierre Dutilleul}, {and} \bibinfo{person}{Donald~L Smith}.} \bibinfo{year}{1999}\natexlab{}.
\newblock \showarticletitle{Advances in the implementation of the box-counting method of fractal dimension estimation}.
\newblock \bibinfo{journal}{\emph{Applied mathematics and computation}} \bibinfo{volume}{105}, \bibinfo{number}{2-3} (\bibinfo{year}{1999}), \bibinfo{pages}{195--210}.
\newblock


\bibitem[Gbur(2011)]%
        {infinityMirror}
\bibfield{author}{\bibinfo{person}{Greg Gbur}.} \bibinfo{year}{2011}\natexlab{}.
\newblock \bibinfo{title}{Infinity is weird, even in infinity mirrors}.
\newblock
\urldef\tempurl%
\url{https://skullsinthestars.com/2011/07/30/infinity-is-weird-even-in-infinity-mirrors/}
\showURL{%
\tempurl}
\newblock
\shownote{Accessed: 2026-01-04}.


\bibitem[Kawaharada(2014a)]%
        {Kawaharada}
\bibfield{author}{\bibinfo{person}{Akane Kawaharada}.} \bibinfo{year}{2014}\natexlab{a}.
\newblock \showarticletitle{Fractal patterns created by Ulam's cellular automaton}. In \bibinfo{booktitle}{\emph{2014 Second International Symposium on Computing and Networking}}. IEEE, \bibinfo{pages}{484--486}.
\newblock


\bibitem[Kawaharada(2014b)]%
        {Kawaharada_rule150}
\bibfield{author}{\bibinfo{person}{Akane Kawaharada}.} \bibinfo{year}{2014}\natexlab{b}.
\newblock \showarticletitle{Ulam's cellular automaton and Rule 150}.
\newblock \bibinfo{journal}{\emph{Hokkaido Mathematical Journal}} \bibinfo{volume}{43}, \bibinfo{number}{3} (\bibinfo{year}{2014}), \bibinfo{pages}{361--383}.
\newblock


\bibitem[Last(2025)]%
        {circular_CA}
\bibfield{author}{\bibinfo{person}{First Last}.} \bibinfo{year}{2025}\natexlab{}.
\newblock \showarticletitle{Cellular Automata}.
\newblock In \bibinfo{booktitle}{\emph{Encyclopedia of Geography}}, \bibfield{editor}{\bibinfo{person}{Barney Warf}} (Ed.). \bibinfo{publisher}{SAGE Publications}, \bibinfo{address}{Thousand Oaks, CA}.
\newblock


\bibitem[Liebovitch and Toth(1989)]%
        {BoxCounting}
\bibfield{author}{\bibinfo{person}{Larry~S Liebovitch} {and} \bibinfo{person}{Tibor Toth}.} \bibinfo{year}{1989}\natexlab{}.
\newblock \showarticletitle{A fast algorithm to determine fractal dimensions by box counting}.
\newblock \bibinfo{journal}{\emph{physics Letters A}} \bibinfo{volume}{141}, \bibinfo{number}{8-9} (\bibinfo{year}{1989}), \bibinfo{pages}{386--390}.
\newblock


\bibitem[Okura(2022)]%
        {Okura_tree}
\bibfield{author}{\bibinfo{person}{Fumio Okura}.} \bibinfo{year}{2022}\natexlab{}.
\newblock \showarticletitle{3D modeling and reconstruction of plants and trees: A cross-cutting review across computer graphics, vision, and plant phenotyping}.
\newblock \bibinfo{journal}{\emph{Breeding Science}} \bibinfo{volume}{72}, \bibinfo{number}{1} (\bibinfo{year}{2022}), \bibinfo{pages}{31--47}.
\newblock


\bibitem[Packard and Wolfram(1985)]%
        {Wolfram_2D}
\bibfield{author}{\bibinfo{person}{Norman~H Packard} {and} \bibinfo{person}{Stephen Wolfram}.} \bibinfo{year}{1985}\natexlab{}.
\newblock \showarticletitle{Two-dimensional cellular automata}.
\newblock \bibinfo{journal}{\emph{Journal of Statistical physics}} \bibinfo{volume}{38}, \bibinfo{number}{5} (\bibinfo{year}{1985}), \bibinfo{pages}{901--946}.
\newblock


\bibitem[Palubicki et~al\mbox{.}(2009)]%
        {Palubicki_tree}
\bibfield{author}{\bibinfo{person}{Wojciech Palubicki}, \bibinfo{person}{Kipp Horel}, \bibinfo{person}{Steven Longay}, \bibinfo{person}{Adam Runions}, \bibinfo{person}{Brendan Lane}, \bibinfo{person}{Radom{\'\i}r M{\v{e}}ch}, {and} \bibinfo{person}{Przemyslaw Prusinkiewicz}.} \bibinfo{year}{2009}\natexlab{}.
\newblock \showarticletitle{Self-organizing tree models for image synthesis}.
\newblock \bibinfo{journal}{\emph{ACM Transactions On Graphics (TOG)}} \bibinfo{volume}{28}, \bibinfo{number}{3} (\bibinfo{year}{2009}), \bibinfo{pages}{1--10}.
\newblock


\bibitem[Peitgen et~al\mbox{.}(2004)]%
        {fractal_book_Peitgen}
\bibfield{author}{\bibinfo{person}{Heinz-Otto Peitgen}, \bibinfo{person}{Hartmut J{\"u}rgens}, \bibinfo{person}{Dietmar Saupe}, {and} \bibinfo{person}{Mitchell~J Feigenbaum}.} \bibinfo{year}{2004}\natexlab{}.
\newblock \bibinfo{booktitle}{\emph{Chaos and fractals: new frontiers of science}}. Vol.~\bibinfo{volume}{106}.
\newblock \bibinfo{publisher}{Springer}.
\newblock


\bibitem[Reinhardt(2012)]%
        {mise}
\bibfield{author}{\bibinfo{person}{Dagmar Reinhardt}.} \bibinfo{year}{2012}\natexlab{}.
\newblock \bibinfo{booktitle}{\emph{Youtopia. a Passion for the Dark: Architecture at the Intersection Between Digital Processes and Theatrical Performance}}.
\newblock \bibinfo{publisher}{Freerange Press}.
\newblock


\bibitem[Rian et~al\mbox{.}(2007)]%
        {hindu_fractal}
\bibfield{author}{\bibinfo{person}{Iasef~Md Rian}, \bibinfo{person}{Jin-Ho Park}, \bibinfo{person}{Hyung~Uk Ahn}, {and} \bibinfo{person}{Dongkuk Chang}.} \bibinfo{year}{2007}\natexlab{}.
\newblock \showarticletitle{Fractal geometry as the synthesis of Hindu cosmology in Kandariya Mahadev temple, Khajuraho}.
\newblock \bibinfo{journal}{\emph{Building and Environment}} \bibinfo{volume}{42}, \bibinfo{number}{12} (\bibinfo{year}{2007}), \bibinfo{pages}{4093--4107}.
\newblock


\bibitem[Sarkar and Chaudhuri(1994)]%
        {Sarkar_DBC}
\bibfield{author}{\bibinfo{person}{Nirupam Sarkar} {and} \bibinfo{person}{Bidyut~Baran Chaudhuri}.} \bibinfo{year}{1994}\natexlab{}.
\newblock \showarticletitle{An efficient differential box-counting approach to compute fractal dimension of image}.
\newblock \bibinfo{journal}{\emph{IEEE Transactions on systems, man, and cybernetics}} \bibinfo{volume}{24}, \bibinfo{number}{1} (\bibinfo{year}{1994}), \bibinfo{pages}{115--120}.
\newblock


\bibitem[Schiff(2011)]%
        {Schiff_CA}
\bibfield{author}{\bibinfo{person}{Joel~L Schiff}.} \bibinfo{year}{2011}\natexlab{}.
\newblock \bibinfo{booktitle}{\emph{Cellular automata: a discrete view of the world}}.
\newblock \bibinfo{publisher}{John Wiley \& Sons}.
\newblock


\bibitem[Stewart(2006)]%
        {pascalTriangle_mod2}
\bibfield{author}{\bibinfo{person}{Ian Stewart}.} \bibinfo{year}{2006}\natexlab{}.
\newblock \bibinfo{booktitle}{\emph{How to cut a cake: and other mathematical conundrums}}.
\newblock \bibinfo{publisher}{OUP Oxford}.
\newblock


\bibitem[{The Divine India}(2024)]%
        {lila}
\bibfield{author}{\bibinfo{person}{{The Divine India}}.} \bibinfo{year}{2024}\natexlab{}.
\newblock \bibinfo{title}{Beautiful Ceiling Design at Akshardham Temple}.
\newblock
\newblock
\shownote{\url{https://www.thedivineindia.com/beautiful-ceiling-design-at-akshardham-temple/image/85}}.


\bibitem[Trivedi(1989)]%
        {hindu_fractal2}
\bibfield{author}{\bibinfo{person}{Kirti Trivedi}.} \bibinfo{year}{1989}\natexlab{}.
\newblock \showarticletitle{Hindu temples: Models of a fractal universe}.
\newblock \bibinfo{journal}{\emph{The Visual Computer}} \bibinfo{volume}{5}, \bibinfo{number}{4} (\bibinfo{year}{1989}), \bibinfo{pages}{243--258}.
\newblock


\bibitem[Ulam et~al\mbox{.}(1962)]%
        {Ulam_uwca}
\bibfield{author}{\bibinfo{person}{Stanislaw Ulam} {et~al\mbox{.}}} \bibinfo{year}{1962}\natexlab{}.
\newblock \showarticletitle{On some mathematical problems connected with patterns of growth of figures}. In \bibinfo{booktitle}{\emph{Proceedings of symposia in applied mathematics}}, Vol.~\bibinfo{volume}{14}. American Mathematical Society Providence, RI, USA, \bibinfo{pages}{215--224}.
\newblock


\bibitem[Watling(2009)]%
        {Watling_Mise}
\bibfield{author}{\bibinfo{person}{Stuart Watling}.} \bibinfo{year}{2009}\natexlab{}.
\newblock \showarticletitle{Medieval ‘Mise-en-Abyme’: The Object Depicted within Itself}.
\newblock \bibinfo{journal}{\emph{The Courtland Institute of Art, Reino Unido, fev}} (\bibinfo{year}{2009}), \bibinfo{pages}{1--2}.
\newblock


\bibitem[Wolfram and Gad-el Hak(2003)]%
        {Wolfram_new}
\bibfield{author}{\bibinfo{person}{Stephen Wolfram} {and} \bibinfo{person}{M Gad-el Hak}.} \bibinfo{year}{2003}\natexlab{}.
\newblock \showarticletitle{A new kind of science}.
\newblock \bibinfo{journal}{\emph{Appl. Mech. Rev.}} \bibinfo{volume}{56}, \bibinfo{number}{2} (\bibinfo{year}{2003}), \bibinfo{pages}{B18--B19}.
\newblock


\bibitem[Zekrizadeh et~al\mbox{.}(2019)]%
        {cole}
\bibfield{author}{\bibinfo{person}{Neda Zekrizadeh}, \bibinfo{person}{Ahmad Khademzadeh}, {and} \bibinfo{person}{Mehdi Hosseinzadeh}.} \bibinfo{year}{2019}\natexlab{}.
\newblock \showarticletitle{An online cost-based job scheduling method by cellular automata in cloud computing environment}.
\newblock \bibinfo{journal}{\emph{Wireless Personal Communications}} \bibinfo{volume}{105}, \bibinfo{number}{3} (\bibinfo{year}{2019}), \bibinfo{pages}{913--939}.
\newblock


\end{thebibliography}

\end{document}